\documentclass[english]{article}
\usepackage[affil-it]{authblk}

\usepackage[T1]{fontenc}
\usepackage[latin9]{inputenc}
\usepackage{refstyle}
\usepackage{color,hyperref}
    \catcode`\_=11\relax
    \newcommand\email[1]{\_email #1\q_nil}
    \def\_email#1@#2\q_nil{%
      \href{mailto:#1@#2}{{\emailfont #1\emailampersat #2}}
    }
    \newcommand\emailfont{\sffamily}
    \newcommand\emailampersat{{\color{red}\small@}}
    \catcode`\_=8\relax

\makeatletter

\RS@ifundefined{subref}
  {\def\RSsubtxt{section~}\newref{sub}{name = \RSsubtxt}}
  {}
\RS@ifundefined{thmref}
  {\def\RSthmtxt{theorem~}\newref{thm}{name = \RSthmtxt}}
  
  {}
\RS@ifundefined{lemref}
  {\def\RSlemtxt{lemma~}\newref{lem}{name = \RSlemtxt}}
  {}

\usepackage{amsmath}

\makeatother

\usepackage{babel}
\usepackage{mathrsfs}
\begin{document}
\title{Projective representation of the Galilei group for classical and quantum-classical systems}
\author{A. D. Berm\'udez Manjarres}
\affil{\footnotesize Universidad Distrital Francisco Jos\'e de Caldas\\ Cra 7 No. 40B-53, Bogot\'a, Colombia\\ \email{ad.bermudez168@uniandes.edu.co}}

\maketitle
\begin{abstract}
A physically relevant unitary irreducible non-projective representation
of the Galilei group is possible in the Koopman-von Neumann formulation
of classical mechanics. This classical representation is characterized
by the vanishing of the central charge of the Galilei algebra. This
is in contrast to the quantum case where the mass plays the role of
the central charge. Here we show, by direct construction, that classical
mechanics also allows for a projective representation of the Galilei
group where the mass is the central charge of the algebra. We extend
the result to certain kind of quantum-classical hybrid systems.
\end{abstract}

\section{Introduction }

It was realized long time ago that classical mechanics can be stated
in the same mathematical formalism of quantum mechanics, i.e., as
a theory of operators acting on a Hilbert space \cite{KvN1,KvN2}.
This operational formalism is known as the Koopman-von Neumann (KvN)
theory.

The KvN theory allowed a classical unitary representation of the Galilei
group given by Loinger \cite{KvN3,KvN4}, while Gulmanelli extended
the results to the Poincar\'e group \cite{KvN5}. A general theory of
Lie group representation in the KvN theory was given by Lugarini and Pauri \cite{KvN6}.
More recently, inspired by the KvN theory (though with a more general
scope), a procedure to obtain classical dynamics as a unitary irreducible
representation of the Galilei group was given in Ref \cite{KvN7},
and the generalization to the relativistic case was considered in
\cite{KvN8}. An interesting aspect of all of the above works is that
the central charge of the Galilei algebra vanishes. This is, contrary
to the quantum case, the classical unitary representations of the Galilei
group considered so far are not projective. However, while the KvN
theory makes unavoidable the vanishing of the central charge, classical
dynamics does not demand it, a possibility that was overlooked. 

Here we consider a variation of the KvN theory recently named as the
Koopman-von Hove theory (KvH) \cite{tronci1,tronci2,tronci3, tronci4}. Although equally
valid at the classical level, the KvH theory puts more emphasis on
the phase of the classical Koopman wavefunctions.

We will show that the KvH leads to a projective representation of
the Galilei group where the mass plays the role of the central charge
of the theory, just as in quantum mechanics. This is, the KvH representation
of the Galilei group is up to a phase.

Besides the purely classical results, the KvN and KvH formalism are
of interest to contrast/compare quantum and classical mechanics and
the study of quantization/dequantization rules \cite{Q1,Q2,Q3,Q4,Q5}.
The KvN and KvH theories are also of interest for the formulation
of some quantum-classical hybrid theories \cite{tronci1,tronci2,sudarshan0,sudarshan1,sudarshan2,sudarshan3,barcelo,hybrid,hybrid2,hybrid3-1}.
In this last regard, the knowledge of the non-relativistic space-time
symmetries of the KvN theory was used to study the viability of quantum-classical
hybrids of the Sudarshan type \cite{hybrid3}. It was shown there
that in Sudarshan hybrids, translational invariance does not equate
to the conservation of the total linear momentum. It was concluded
that Sudarshan hybrids were at odds with momentum conservation. Unfortunately,
we will see that this conclusion was hastily made. We will show that
the hybrid systems recently considered by Bondar, Gay-Balmaz, and
Tronci \cite{tronci1} can be Galilei covariant and conserve the total
momentum. 

This work is organized as follows: in section 2 we review the KvN
and KvH Hilbert space formulations of classical mechanics. We will
focus on the construction of Hermitian operators from phase space
functions.

In section 3 the elements of the Galilei algebra will be related to
Hermitian operators constructed from the KvN and KvH. It is shown
there that the KvH theory leads to a projective representation of
the Galilei group.

In section 4 we analyze the Galilean covariance and momentum conservation
of quantum-classical hybrid systems of the Sudarshan type obtained
from the KvH formalism. We will see that the KvH theory leads naturally
to a hybrid interaction term that conserves the total momentum and
is not at odds with Galilean covariance.

\section{The KvN and KvH mechanics}

It is possible to associate a Hilbert $\mathcal{H}_{cl}$ space to
a given phase space $\varGamma$ . The procedure is as follows:
let $d\omega$ be the usual symplectic measure of $\varGamma$, the
vectors of $\mathcal{H}_{cl}$ are then given by square integrable
functions 

\begin{equation}
\int_{\varGamma}\left|\psi\right|^{2}d\omega<\infty.
\end{equation}
The inner product on $\mathcal{H}_{cl}$ can be defined as
\begin{equation}
\left\langle \psi_{1}\right.\left|\psi_{2}\right\rangle =\int_{\varGamma}\psi_{1}^{*}\psi_{2}d\omega.\label{inner}
\end{equation}
From now on, we will specialize for the case of a single particle
moving in $R^{3}$, such that $d\omega=d\mathbf{q}d\mathbf{p}$
in Cartesian coordinates. 

There are infinite ways to use phase space functions to define linear
operators acting on $\mathcal{H}_{cl}$ . The easiest method to define
a Hermitian operator from a space space function is by multiplication
as follows:

\begin{equation}
\hat{f}(\mathbf{q},\mathbf{p})\psi(\mathbf{q},\mathbf{p})=f(\mathbf{q},\mathbf{p})\psi(\mathbf{q},\mathbf{p}).
\end{equation}
The most important operators defined this way are the position and
momentum operators 
\begin{align}
\hat{\mathbf{q}}f(\mathbf{q},\mathbf{p}) & =\mathbf{q}f(\mathbf{q},\mathbf{p}),\nonumber \\
\hat{\mathbf{p}}f(\mathbf{q},\mathbf{p}) & =\mathbf{p}f(\mathbf{q},\mathbf{p}).
\end{align}
Let us note that these so-defined position and momentum operators
commute with each other $\left[q_{i},p_{j}\right]=0$; they do not
have any uncertainty principle associated to them. Thus, they can
be simultaneously measured, as they should in any classical theory.
We also point out that $\hat{\mathbf{q}}$ and $\hat{\mathbf{p}}$
form a complete set of commuting observables since we are dealing
with a single structureless particle.

There are other ways to define Hermitian operators from phase space
functions. A rather natural way comes from using the Poisson bracket
as follows:
\begin{equation}
\hat{F}=-i\left\{ \cdot,f(\mathbf{q},\mathbf{p})\right\} .\label{kvNBracket}
\end{equation}
We will call the procedure given by Eq.(\ref{kvNBracket}) as the
KvN rule. Using the the KvN rule on $(\mathbf{q},\mathbf{p})$ we
get

\begin{align}
\widehat{\lambda}{}_{\mathbf{q}} & =-i\left\{ \,,\mathbf{p}\right\} =-i\nabla_{\mathbf{q}},\nonumber \\
\widehat{\lambda}_{\mathbf{p}} & =i\left\{ \,,\mathbf{q}\right\} =-i\nabla_{\mathbf{p}}.
\end{align}
From their definition, we can write the following set of commutation
relations

\begin{eqnarray}
\left[\widehat{q}_{i},\widehat{q}_{_{j}}\right] & = & \left[\widehat{q}_{i},\widehat{p}_{j}\right]=\left[\widehat{p}_{i},\widehat{p}_{j}\right]=0,\nonumber \\
\left[\widehat{q}_{i},\widehat{\lambda}_{p_{i}}\right] & = & \left[\widehat{p}_{i},\widehat{\lambda}{}_{q_{j}}\right]=\left[\widehat{\lambda}{}_{q_{j}},\widehat{\lambda}_{p_{i}}\right]=0,\nonumber \\
\left[\widehat{q}_{i},\widehat{\lambda}{}_{q_{j}}\right] & = & \left[\widehat{p}_{i},\widehat{\lambda}_{p_{j}}\right]=i\delta_{ij}.\label{Xlambda-1}
\end{eqnarray}
The set $(\hat{\mathbf{q}},\hat{\mathbf{p}},\widehat{\lambda}{}_{\mathbf{q}},\widehat{\lambda}_{\mathbf{p}})$
is irreducible on $\mathcal{H}_{cl}$ in view of commutation relations
(\ref{Xlambda-1}) and Schur lemma.

Acting on the Hamiltonian function $H=\frac{\mathbf{p}^{2}}{2m}+V(\mathbf{q})$,
the KvN rule gives the following operator

\begin{equation}
\widehat{L}=-i\left\{ ,H\right\} =\frac{1}{m}\hat{\mathbf{p}}\cdot\hat{\lambda}_{\mathbf{q}}-\nabla_{\mathbf{q}}V(\mathbf{q})\cdot\hat{\lambda}_{\mathbf{p}}.\label{Liouvillian}
\end{equation}
The operator $\widehat{L}$ is known as the Liouvillian, and it is
of central importance since it governs time evolution. Consider the
Liouville equation of classical statistical mechanics

\begin{equation}
\frac{\partial\rho}{\partial t}=-\left\{ \rho,H\right\} =-i\widehat{L}\rho.\label{L-1}
\end{equation}
Due to the linearity in $\hat{\lambda}_{\mathbf{q}}$ and $\hat{\lambda}_{\mathbf{p}}$
of (\ref{L-1}), the square integrable function defined by \footnote{A constant in needed here to make the exponent dimensionless. This
constant is usually chosen to be $\hbar^{-1}$. In this work we set
$\hbar$=1.} $\psi=\sqrt{\rho}e^{iS}$ also obeys the Liouville equation
\begin{equation}
\frac{\partial\psi}{\partial t}=-i\widehat{L}\psi.\label{sch}
\end{equation}
The equation (\ref{sch}) has the form of a Schr\"odinger equation,
where $\widehat{L}$ takes the role of the ``Hamiltonian'' operator
of the theory. Hence, classical statistical mechanics is equivalent
to solving the Schr\"odinger-like equation (\ref{sch}). Particle mechanics
and Hamilton equations are formally recovered with the use Klimontovich
delta distributions, i.e., $\rho=\delta(\mathbf{p}(t),\mathbf{q}(t))$.

Notice that in the KvN theory the modulus and the phase of $\psi$ are decoupled. Replacing $\psi=\sqrt{\rho}e^{iS}$ into (\ref{sch}) leads to 

\begin{align}
\frac{\partial\rho}{\partial t}+\left\{ \rho,H\right\} 	& =0, \nonumber \\
\frac{\partial S}{\partial t}+\left\{ S,H\right\} 	& =0.
\end{align}

As a consequence, the KvN evolution is such that $\psi(without\,phase\,at\,t=0)\rightarrow\psi(without\,phase\,at\,t)$. So, contrary to what happens in quantum mechanics, KvN phases cannot be generated during time evolution \cite{Mauro}, and $\psi$ can be simply chosen to be real-valued.

Another viable formula for assigning Hermitian operators to phase
space functions is 
\begin{equation}
\hat{F}^{\star}=-i\left\{ \cdot,f\right\} +f-\mathbf{p}\cdot\frac{\partial f}{\partial\mathbf{p}}.\label{KvHrule}
\end{equation}
After the convention used in \cite{tronci1,tronci2}, we call the
association (\ref{KvHrule}) as the KvH rule\footnote{There is a gauge freedom in the assignation of the Hermitian operator
given by the KvH rule, see Ref \cite{tronci1}, but we will stick
with the equation (\ref{KvHrule}) for the rest of this work. }. Applying the KvH rule on the Hamiltonian we get
\begin{equation}
\widehat{L}^{\star}=-i\left\{ \cdot,H\right\} +H-\mathbf{p}\cdot\frac{\partial H}{\partial\mathbf{p}}=\widehat{L}-\mathscr{L},\label{KvHL}
\end{equation}
where $\widehat{L}$ is the operator given by (\ref{Liouvillian}),
and $\mathscr{L}=\mathbf{p}\cdot\frac{\partial H}{\partial\mathbf{p}}-H$
is the Lagrangian associated to $H$.

Acting on $\psi=\sqrt{\rho}e^{iS}$, the Schr\"odinger-like equation
generated by $\widehat{L}^{\star}$ 
\begin{equation}
\frac{\partial\psi}{\partial t}=-i\widehat{L}^{\star}\psi,\label{sch-1}
\end{equation}
splits into

\begin{align}
\frac{\partial\rho}{\partial t}+\left\{ \rho,H\right\}  & =0,\nonumber \\
\frac{\partial S}{\partial t}+\left\{ S,H\right\}  & =\mathscr{L}.
\end{align}
We can formally solve for the phase as $S=\int_{\eta}\mathscr{L}dt+constant$, where $\eta$ means that the integral has to be evaluated over the phase-space flow produced by $\left\{ ,H\right\}$. Hence, we can see that $S$ can be identified with the classical action of analytical mechanics \cite{Q4, action1}. Thus, the KvH dynamics is equivalent to the classical Liouville equation, but, contrary to the KvN case, the KvH equation also contains information about the classical phase (see \cite{action2, action3}  for further information about the relation between modulus and phase in Koopman wavefunctions). 

It is worth mentioning an important third possibility to associate
Hermitian operators to face space functions, although we will not
use it in this work. Instead of the Poisson bracket, the star product
of Wigner phase-space representation of quantum mechanics can be used
\cite{star}. The use of the star product leads to quantum representation
of the Galilei group.

\section{The KvN and KvH Representations of the Galilei group}

In Hamiltonian mechanics, the space-time transformations associated
with the Galilei group are realized via canonical transformations.
The generator functions associated with these canonical transformations
are the linear momentum $\mathbf{p},$ the angular momentum $\mathbf{j}=\mathbf{r}\times\mathbf{p}$,
the dynamic mass moment $\mathbf{g}=m\mathbf{q}-t\mathbf{p}$ and
the Hamiltonian of the free particle $H_{free}=\frac{\mathbf{p}^{2}}{2m}$.
They realize the Lie algebra of the Galilei group via the Poisson
bracket. This is known as a canonical representation of a Lie group
\cite{canonical,canonical2}. 

The canonical representation of the Galilei group can be reinterpreted
as a unitary representation in the context of the KvN theory as follows:
first, the operator $\widehat{\lambda}{}_{\mathbf{q}}$ is the generator
of translation. Its action on an element of $\mathcal{H}(\Gamma)$
is given by
\begin{equation}
\exp\left[-i\mathbf{a}\cdot\widehat{\lambda}{}_{\mathbf{q}}\right]\psi(\mathbf{q},\mathbf{p})=\psi(\mathbf{q}-\mathbf{a},\mathbf{p}).\label{translation}
\end{equation}
On the other hand, while $\hat{\lambda}_{\mathbf{p}}$ translate the
momentum coordinates, \begin{equation}\exp\left[-i\mathbf{b}\cdot\widehat{\lambda}{}_{\mathbf{p}}\right]\psi(\mathbf{q},\mathbf{p})=\psi(\mathbf{q},\mathbf{p}-\mathbf{b}),\end{equation}
$\hat{\lambda}_{\mathbf{p}}$ is not an element of the Galilei algebra. 

The other generators of the Galilei algebra can be obtained with the
KvN rule as
\begin{align}
\mathcal{\hat{J}}_{i} & =-i\left\{ ,j_{i}\right\} =\varepsilon_{ijk}\left(\hat{q}_{j}\hat{\lambda}_{q_{k}}+\hat{p}_{j}\hat{\lambda}_{p_{k}}\right),\nonumber \\
\hat{\mathcal{G}} & =-i\left\{ ,\mathbf{g}\right\} =-t\hat{\lambda}_{\mathbf{q}}-m\hat{\lambda}_{\mathbf{p}},\nonumber \\
\hat{L} & =-i\left\{ ,H_{free}\right\} =\frac{1}{m}\hat{\mathbf{p}}\cdot\hat{\lambda}_{\mathbf{q}}.\label{cla rep}
\end{align}
In the KvN theory $\mathcal{\hat{J}}$ is the generator of rotations,
$\hat{\mathcal{G}}$ is the generator of Galilean boosts, and $\hat{L}$
is the time translation operator for the case of a free particle.
The above operators obey the commutation relations of the Galilei
algebra.

\begin{align}
\left[\widehat{\lambda}_{q_{i}},\widehat{\lambda}_{q_{j}}\right] & =\left[\widehat{\mathcal{G}}_{i},\widehat{\mathcal{G}}_{j}\right]=\left[\widehat{\mathcal{J}}_{i},\widehat{L}\right]=\left[\widehat{\lambda}_{q_{i}},\widehat{L}\right]=0;\nonumber \\
\left[\widehat{\mathcal{J}}_{i},\widehat{\mathcal{J}}_{j}\right] & =i\varepsilon_{ijk}\widehat{\mathcal{J}}_{k};\;\left[\widehat{\mathcal{J}}_{i},\widehat{\lambda}_{q_{k}}\right]=i\varepsilon_{ijk}\widehat{\lambda}_{q_{k}};\nonumber \\
\left[\widehat{\mathcal{J}}_{i},\widehat{\mathcal{G}}_{j}\right] & =i\varepsilon_{ijk}\widehat{\mathcal{G}}_{k};\:\left[\widehat{\mathcal{G}}_{i},\widehat{L}\right]=i\widehat{\lambda}_{q_{i}},\label{galileialgebra}
\end{align}
and

\begin{equation}
\left[\widehat{\mathcal{G}}_{j},\widehat{\lambda}_{q_{i}}\right]=0.\label{nocentralcharge}
\end{equation}

Equations (\ref{galileialgebra}) and (\ref{nocentralcharge}) are
a realization of the Galilei algebra in the very particular case of
a vanishing central charge, as evidenced by the vanishing right-hand-side
of (\ref{nocentralcharge}). Since the set $\left\{ \widehat{\mathbf{q}},\widehat{\mathbf{p}},\widehat{\lambda}{}_{\mathbf{q}},\widehat{\lambda}_{\mathbf{p}}\right\} $
is irreducible in $\mathcal{H}(\Gamma)$, and taking into account
the vanishing of the central charge, the above constitutes a unitary
irreducible faithful representation of the Galilei group.

The remaining space-time transformations of the Galilei group are
realized as unitary transformation given by 

\begin{align}
\exp\left[-i\theta\hat{\mathbf{n}}\cdot\widehat{\mathbf{\mathcal{J}}}\right]\psi(\mathbf{q},\mathbf{p}) & =\psi(\mathbf{q}-\theta\hat{\mathbf{n}}\times\mathbf{q},\mathbf{p}-\theta\hat{\mathbf{n}}\times\mathbf{p}),\nonumber \\
\exp\left[i\mathbf{v}\cdot\hat{\mathcal{G}}\right]\psi(\mathbf{q},\mathbf{p}) & =\psi(\mathbf{q}-\mathbf{v}t,\mathbf{p}-m\mathbf{v}),\nonumber \\
\exp\left[-it\hat{L}\right]\psi(\mathbf{q},\mathbf{p}) & =\psi(\mathbf{q}-\frac{t}{m}\mathbf{p},\mathbf{p}).\label{spacetime}
\end{align}

We now proceed to investigate the effect of the KvH rules on the representation
theory of the Galilei group. The KvH Liouvillian in the free particle
case is given by

\begin{equation}
\hat{L}^{\star}=-i\left\{ ,H_{free}\right\} -H_{free}=\frac{\mathbf{\widehat{p}}}{m}\cdot\hat{\lambda}_{\mathbf{q}}-\frac{\widehat{\mathbf{p}}^{2}}{2m}.\label{L*}
\end{equation}
The boost operator is 
\begin{equation}
\widehat{\mathcal{G}}^{\star}=-i\left\{ ,\mathbf{g}\right\} +\mathbf{g}-\mathbf{p}\cdot\frac{\partial\mathbf{g}}{\partial\mathbf{p}}=-t\hat{\lambda}_{\mathbf{q}}-m\hat{\lambda}_{\mathbf{p}}+m\mathbf{q}.\label{G*}
\end{equation}
All the other generators are unaltered compared to the KvN case. It
can be checked that the replacement $\hat{L}\rightarrow\hat{L}^{\star}$
and $\hat{\mathcal{G}}\rightarrow\hat{\mathcal{G}}^{\star}$ gives
a realization of the Galilei algebra. However, contrary to the KvN
case, now the central charge of the algebra does not vanish. Instead,
we have the commutation relation

\[
\left[\widehat{\mathcal{G}}_{j}^{\star},\widehat{\lambda}_{q_{i}}\right]=im\delta_{ij}.
\]
We can see that the mass is the central charge in the KvH theory,
just as in quantum mechanics. Hence, the unitary representation of
the Galilei group given by the KvH theory is projective.

In the KvH theory, the unitary transformation associated to space
translation and rotations are the same as in Eqs.(\ref{translation})
and (\ref{spacetime}). Galilean boosts and time translation are modified
to include a phase factor as follows
\begin{align}
\exp\left[i\mathbf{v}\cdot\hat{\mathcal{G}^{\star}}\right]\psi(\mathbf{q},\mathbf{p})&=e^{i(m\mathbf{q}\cdot\mathbf{v}-\frac{mt}{2}v^{2})}\psi(\mathbf{q}-\mathbf{v}t,\mathbf{p}-m\mathbf{v}),\\
\exp\left[-it\hat{L}^{\star}\right]\psi(\mathbf{q},\mathbf{p})&=e^{-i\frac{\mathbf{p}^{2}}{2m}t}\psi(\mathbf{q}-\frac{t}{m}\mathbf{p},\mathbf{p}).
\end{align}

Finally, we mention that while $\hat{\lambda}_{\mathbf{q}}$ is the
same in both theories, $\widehat{\lambda}{}_{\mathbf{p}}$ is not.
In the KvH theory we have $\widehat{\lambda}{}_{\mathbf{p}}^{\star}=\widehat{\lambda}{}_{\mathbf{p}}+\mathbf{q}$
and $[\hat{\lambda}_{qi}^{\star},\widehat{\lambda}{}_{pj}^{\star}]=-i\delta_{ij}$.

\subsection{Covariance of composite systems.}

Let us now consider the case of two interacting particles. We can
to construct a representation of the Galilei group on the space $\mathcal{H}_{T}=\mathcal{H}_{cl}^{(1)}\mathrm{\otimes}\mathcal{H}_{cl}^{(2)}$
starting from the representation of the individual sectors. Except
for the Liouvillian, the generators of the algebra acting on $\mathcal{H}_{T}$
are given by the sum of the individual generators; for example, the
total translation operator is given by $\hat{\lambda}_{\mathbf{q}\,total}=\hat{\lambda}_{\mathbf{q}_{1}}+\hat{\lambda}_{\mathbf{q}_{2}}$.
All the commutation relations of the Gailiei algebra not involving
the Liouvillian are satisfied identically. On the other hand, the
total Liouvillian is allowed to have an extra term representing the
interaction between the particles. The allowed interaction term can
then be deduced from considerations of Galilean covariance. Using
the KvH rule on the Hamiltonian 

\[
H=\frac{\mathbf{p}_{1}^{2}}{2m_{1}}+\frac{\mathbf{p}_{2}^{2}}{2m_{2}}+V(\mathbf{q}_{1},\mathbf{q}_{2})
\]
gives the Liouvillian

\begin{align}
\hat{L}^{\star} & =\frac{1}{m_{1}}\hat{\mathbf{p}}_{1}\cdot\hat{\lambda}_{\mathbf{q}_{1}}+\frac{1}{m_{2}}\hat{\mathbf{p}}_{2}\cdot\hat{\lambda}_{\mathbf{q}_{2}}\nonumber \\
 & -\nabla_{\mathbf{q}_{1}}V\cdot\hat{\lambda}_{\mathbf{p}_{1}}-\nabla_{\mathbf{q}_{2}}V\cdot\hat{\lambda}_{\mathbf{p}_{2}}\nonumber \\
 & -\frac{\hat{\mathbf{p}}_{1}^{2}}{2m_{1}}-\frac{\hat{\mathbf{p}}_{2}^{2}}{2m_{2}}+V.\label{2L}
\end{align}

Translational invariance dictates that 

\begin{equation}
\left[\hat{\lambda}_{\mathbf{q}\,total},\hat{L}^{\star}\right]=0.\label{total translation}
\end{equation}
The necessary and sufficient condition for Eq (\ref{total translation})
to be true is that the potential energy depend only on the relative
position $V=V(\mathbf{\hat{q}}_{1}-\hat{\mathbf{q}}_{2})$. The above
also takes care of the commutation relation with the Galilean boost

\begin{equation}
\left[\left(\hat{\mathcal{G}}_{total}^{\star}\right)_{j},\hat{L}^{\star}\right]=\hat{\lambda}_{q_{1}j}+\hat{\lambda}_{q_{2}j}.
\end{equation}
Finally, if $V$ is a scalar then
\begin{equation}
\left[\widehat{\mathcal{J}}_{total},\widehat{L}\right]=0.
\end{equation}

Notice that the condition of translational invariance is not the same
as conservation of the total momentum $\hat{\mathbf{p}}_{total}=\hat{\mathbf{p}}_{1}+\hat{\mathbf{p}}_{2}$.
However, it is true that $V=V(\mathbf{\hat{q}}_{1}-\hat{\mathbf{q}}_{2})$
leads to 
\begin{equation}
\left[\hat{\mathbf{p}}_{total},\hat{L}^{\star}\right]=0.\label{total translation-1}
\end{equation}

\section{Quantum-classical hybrids}

It is possible to obtain the two-body Schr\"odinger equation from the
Liouvillian (\ref{2L}). Following Klein \cite{Q4}, the procedure
consist of two steps: (1) impose the condition $\partial_{p}\psi=0$
for every momentum coordinate, (2) make the replacement 
$p\rightarrow-i\partial_{q}$, this correspond to the usual canonical
quantization rule. For the Liouvillian (\ref{2L}), the result is
\begin{equation}
i\frac{\partial\psi}{\partial t}=-\frac{1}{2m_{1}}\nabla_{\mathbf{q}1}^{2}\psi-\frac{1}{2m_{2}}\nabla_{\mathbf{q}2}^{2}\psi+V\psi.
\end{equation}

Instead of doing a full quantization of (\ref{2L}), we can obtain
a hybrid theory by applying Klein's method to only one particle. This
will result in a special case of an hybrid theory of the Sudarshan
type \cite{tronci1,tronci2,sudarshan0}. To distinguish the classical
degrees of freedom from the quantum ones, we keep the $(\mathbf{q},\mathbf{p})$
notation for the classical particle and use $\mathbf{x}$ and $\mathbf{k}$
for the quantum coordinate and momentum, respectively. The quantum
position and momentum operators obeys the usual Heisenberg commutation
relations $\left[\hat{x}_{i},\hat{k}_{j}\right]=i\delta_{ij}$. The
hybrid Hilbert space is of the form $\mathcal{H}_{T}=\mathcal{H}_{\mathrm{Q}}\mathrm{\otimes}\mathcal{H}_{cl}$,
where the vectors of the quantum sector $\mathcal{H}_{\mathrm{Q}}$
are given by normalizable linear combinations of the kets $\left|\mathbf{x}\right\rangle $.

In the quantum sector, and for a single spinless particle, the elements
of the Galilei algebra in terms $\mathbf{x}$ of $\mathbf{k}$ are:
$\mathbf{k}$ itself, the angular momentum $\hat{\mathbf{j}}=\hat{\mathbf{x}}\times\hat{\mathbf{k}}$,
the dynamic mass moment $\hat{\mathbf{g}}=m\hat{\mathbf{x}}-t\hat{\mathbf{k}}$,
and the Hamiltonian $\hat{H}_{\mathrm{Q}}=\frac{\hat{\mathbf{k}}^{2}}{2m}$,
where $m$ is the central charge of the algebra \cite{qgalilei ,qgalilei2,qgalilei3,qgalilei4}.
In the quantum case, $\mathbf{k}$ has the dual role of being the
momentum operator and the generator of spatial translations.

Performing the partial quantization of (\ref{2L}) results in the
hybrid total time-translation operator 

\begin{align}
\hat{L}_{h} & =\frac{1}{m_{1}}\hat{\mathbf{p}}\cdot\hat{\lambda}_{\mathbf{q}}-\frac{\hat{\mathbf{p}}^{2}}{2m_{1}}+\frac{1}{2m_{2}}\hat{\mathbf{k}}^{2}\nonumber \\
 & -\nabla_{\mathbf{q}}V\cdot\hat{\lambda}_{\mathbf{p}}+V.\label{2L-2}
\end{align}

As in section 3.1, the commutation relations of the Galilei algebra
of the composite systems not involving the time translation operator
are all satisfied identically, and the ones that do involve $\hat{L}_{h}$
give conditions for the allowed interaction terms. We emphasize here
that in general Sudarshan hybrids, the conditions for Galilean covariance
do not imply the conservation of quantities like the linear or the
angular momentum. For example, an interaction term can be translational
invariant without conserving the total linear momentum \cite{hybrid3}.
However, we will focus our attention on the particular interaction
term given in (\ref{2L-2}).

A translational invariant potential $V=V(\hat{\mathbf{q}}-\hat{\mathbf{x}})$
guarantees the translational invariance of the hybrid Liouvillian
(\ref{2L-2})

\begin{equation}
\left[\hat{\lambda}_{\mathbf{q}}+\hat{\mathbf{k}},\hat{L}_{h}\right]=0,
\end{equation}
and the conservation of the total momentum
\begin{equation}
\left[\hat{\mathbf{p}}+\hat{\mathbf{k}},\hat{L}_{h}\right]=0.
\end{equation}
Rotational invariance of is obtained if $V$ is a scalar.

With $\hat{\mathcal{G}}_{hybrid}=m_{2}\hat{\mathbf{x}}-t\hat{\mathbf{k}}-t\hat{\lambda}_{\mathbf{q}}-m_{1}\hat{\lambda}_{\mathbf{p}}+m_{1}\mathbf{q}$,
it is immediate to check that the central charge of the algebra is
the total mass
\begin{equation}
\left[\left(\hat{\mathcal{G}}_{hybrid}\right)_{i},\hat{\lambda}_{qj}+\hat{k}_{j}\right]=i\left(m_{1}+m_{2}\right)\delta_{ij}.
\end{equation}
Hence, the hybrid representation of the Galilei group is projective.

It is worth emphasizing that it is the combination $-\nabla_{\mathbf{q}}V\cdot\hat{\lambda}_{\mathbf{p}}+V$
in (\ref{2L-2}) that allows translational invariance and the conservation
of the linear momentum at the same time. Neither $V$ nor $-\nabla_{\mathbf{q}}V\cdot\hat{\lambda}_{\mathbf{p}}$
do the trick on their own.

\section{Conclusions}

In quantum mechanics, the central charge of the Galilei algebra is
the mass, and it plays the role of a superselection operator. In the
quantum case, representations with a vanishing central charge are
unphysical. We have shown that in the KvH representation of the Galilei
group the mass is the central charge. It is interesting that classical
mechanics allows, but does not require, projective unitary representations
of the Galilei group. At the classical level, whether the central
charge vanishes or not has no effect since the KvN and the KvH formalism
are equally valid.

For hybrid systems, the KvH rule and the Klein quantization method
lead to the specific interaction term $-\nabla_{\mathbf{q}}V\cdot\hat{\lambda}_{\mathbf{p}}+V$
that guarantees Galilean covariance and conservation of the total
momentum.

\end{document}